\documentclass[
    ,final            
  ]
  {aipproc}

\layoutstyle{6x9}

\begin{document}

    \def\farcm{\mbox{\ensuremath{.\mkern-4mu^\prime}}}
    \def\farcs{\mbox{\ensuremath{.\!\!^{\prime\prime}}}}
    \def\etal{${\rm \hspace*{0.8ex}et\hspace*{0.7ex}al.\hspace*{0.6ex}}$}
    \def\plus{${\rm \hspace*{0.7ex}\&\hspace*{0.65ex}}$}
    \def\ie{i.\,e.\ }
    \def\eg{e.\,g.\ }

\title{Gas Evolution in Protoplanetary Disks}


\author{Peter Woitke}{
  address={UK Astronomy Technology Centre, Royal Observatory,
           Blackford Hill, Edinburgh EH9\,3HJ, UK}
}

\author{Bill Dent}{
  address={UK Astronomy Technology Centre, Royal Observatory,
           Blackford Hill, Edinburgh EH9\,3HJ, UK}
}

\author{Wing-Fai Thi}{
  address={UK Astronomy Technology Centre, Royal Observatory,
           Blackford Hill, Edinburgh EH9\,3HJ, UK}
}

\author{Bruce Sibthorpe}{
  address={UK Astronomy Technology Centre, Royal Observatory,
           Blackford Hill, Edinburgh EH9\,3HJ, UK}
}

\author{Ken Rice}{
  address={Institute for Astronomy, Royal Observatory, 
           Blackford Hill, Edinburgh EH9\,3HJ, UK}
}

\author{Jonathan Williams}{
  address={Institute for Astronomy, University of Hawaii,
           2680 Woodlawn Drive, Honolulu HI 96822}
}

\author{Aurora Sicilia-Aguilar}{
  address={Max-Planck-Institut f\"{u}r Astronomie, K\"{o}nigstuhl 17, 
           69117 Heidelberg, Germany}
}

\author{Joanna Brown}{
  address={Max-Planck-Institut f\"{u}r extraterrestrische Physik, 
           Postfach 1312, 85741 Garching, Germany}
}

\author{Inga Kamp}{
  address={Kapteyn Astronomidcal Institute, Postbus 800,
             9700 AV Groningen, The Netherlands}
}

\author{Ilaria Pascucci}{
  address={Steward Observatory, The University of Arizona,
           933 N.~Cherry Avenue, Tucson, AZ - 85718, USA }
}

\author{Richard Alexander}{
  address={Sterrewacht Leiden, Universiteit Leiden, Niels Bohrweg 2, 
           2300 RA Leiden, The Netherlands}
}

\author{Aki Roberge}{
  address={NASA Goddard Space Flight Center,
           Code 667, Greenbelt, MD 20771, USA}
}

\begin{abstract}
This article summarizes a Splinter Session at the Cool Stars XV
conference in St.~Andrews with 3 review and 4 contributed talks. The
speakers have discussed various approaches to understand the structure
and evolution of the gas component in protoplanetary disks. These
ranged from observational spectroscopy in the UV, infrared and
millimeter, through to chemical and hydrodynamical models. 
The focus was on disks around low-mass stars, ranging from classical
T\,Tauri stars to transitional disks and debris disks. Emphasis was
put on water and organic molecules, the relation to planet formation,
and the formation of holes and gaps in the inner regions.

\end{abstract}

\maketitle


\section{Introduction}

Most of the current knowledge about the structure and evolution of
protoplanetary disks (photometry, SED class, images in scattered
light, etc.) originate from the dust component, whereas the gas in
protoplanetary disks is, in many cases, more difficult to observe and
more difficult to model (see Fig.~1). Yet, it is the gas component
that contains 99\% of the disk mass and sets the initial conditions
for planet formation. The gas is responsible for the delivery of
water and organic material to terrestrial planets and, by its
dispersal, drives the evolution to debris disks.

With existing and new telescopes ({\sc Spitzer, Herschel, Sofia, Jwst,
Alma}), tracers of the gas in form of emission lines of atoms, ions
and molecules can be observed. These observations allow for a direct
determination of the gas mass, the temperature distribution and the
chemical composition of the gas independent of the dust, provided that
we can trust the models. By comparing objects of different ages, the
evolution of the gas in protoplanetary disks can be revealed.

Since UV, infrared, and millimeter lines probe very different regions
in the disks, observations generally need to be
combined to obtain a complete and consistent picture of the physical
and chemical state of the gas in protoplanetary disks. Here, models
play an essential role to convert observations into understanding.
Given these new challenges, astronomers and theoreticians around the
world have started to ``hunt'' the gas in protoplanetary disks.  Key
questions are:
\begin{itemize}
\item How does the gas-to-dust ratio vary with age, starting from the 
      initial interstellar value of 100:1 to virtually 0:1 in debris disks?
\item What is the density structure, temperature, and dynamics of
      the gas in protoplanetary disks at different ages?  
\item What is the chemical composition of the gas in the regions
      where terrestrial and where gas giant planets form?
\item What is the driving mechanism for the gas mass loss?  How does the
      dust component react on the gas removal?  
\item Does the gas removal truncate the process of planet formation?
\end{itemize}

\section{Talk Summaries}

\subsubsection{1. Current and Future Observations of Gas Disks}

\begin{figure}
  \begin{minipage}{17cm}
  \center    
  {\ }\\*[-5mm]
  \includegraphics[height=5.4cm,clip=true]{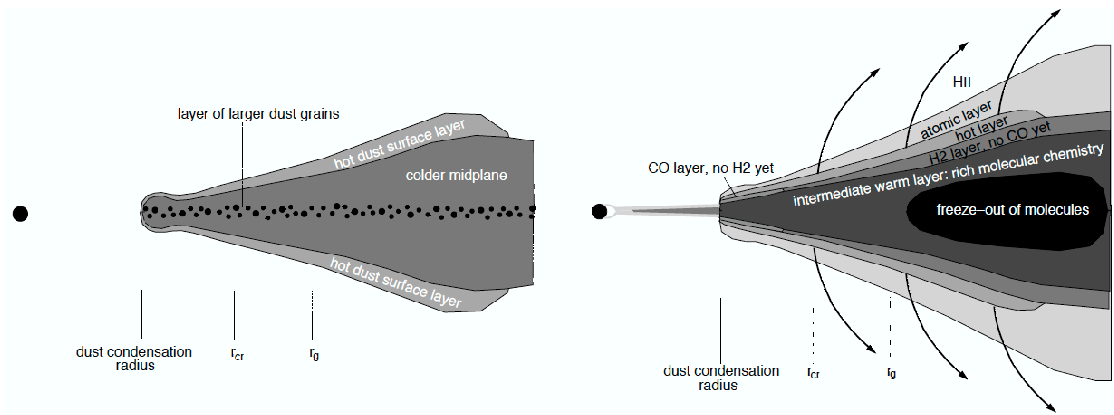} 
  \end{minipage}
  \caption{Comparison of the structure of a flaring protoplanetary
  disk in dust (left) and gas (right). From Dullemond et al.\ (2007).}
  \nocite{Dullemond2007}
\end{figure}

\noindent {\bf Jonathan Williams} reviewed the observational methods
from the UV to millimeter, outlining how spectral profiles can be used
to derive gas properties. Starting from the inner disk, at radii
up to a few AU, near-IR transitions such as the fundamental CO lines
around 4.7$\,\mu$m can probe the regime of terrestrial planets, and
are readily detectable in Classical T\,Tauri stars.  They have been
used to show that gas is present {\em within} the co-rotation and
dust-emitting radii (Carr 2007, Pontoppidan et al.\
2008). \nocite{Carr2007, Pontoppidan2008}

Spitzer mid-infrared spectroscopy has probed gas further out in the
terrestrial zone. The results have set limits on H$_2$ (\eg
Pascucci\etal 2007)\nocite{Pascucci2007}, have shown strong [Ne\,{\sc
ii}] emission from X-ray ionization of the disk surface (\eg Lahuis et
al.\ 2007)\nocite{Lahuis2007}, and detected disks full of water and
organic molecules (Watson\etal 2007, Salyk\etal
2008). \nocite{Watson2007,Salyk2008}

Millimeter lines probe the cold outer reservoir, and CO rotational
lines are detected in $\sim$100 disks. 
However, photo-dissociation and freeze-out on grains can greatly affect
the gas abundances, which makes mass estimates uncertain.  CO is
rarely detectable around the older debris disks (\eg Dent\etal
2005)\nocite{Dent2005} and the results suggest that the gas
dissipates from the inside out (\eg Hughes\etal
2008).\nocite{Hughes2008} One key observation in these disks is the
D/H ratio. This is an important method of constraining the origin of
water on Earth, and the first observational measurements of the D/H
profile across a disk have recently been made by
Qi\etal(2008)\nocite{Qi2008}.

In the future, observations of gas in disks will be greatly advanced
by the spectroscopic capabilities of upcoming missions.  In 2009, the
3m-class telescopes {\sc Sofia} and {\sc Herschel} will open up the
far-infrared wavelength range, looking at gas and dust in the giant
planet formation regions.  By 2012, {\sc Alma} will provide orders of
magnitude improvement in resolution and sensitivity in the sub-mm,
diagnosing gas in the cold disk, and potentially directly imaging
giant proto-planets.  In 2013, {\sc Jwst} will provide similar improvements
in the near and mid-IR. Later in this decade, telescopes
such as TMT and ELT will 
allow diagnosis of gas and dust in the terrestrial planet forming zone. 

By combining data from these new missions with models of the disks, we
hope to understand the gas dissipation and evolution, the structure,
dynamics and chemistry, and the effect and signatures of protoplanets
within these disks. But in addition, we can expect such new facilities
to turn up the unexpected.

\subsubsection{2. Accretion in Evolved and Transitional Protoplanetary Disks}

\begin{figure}[b]
  \begin{minipage}{10cm}
  \center
  {\ }\\*[-5mm]
  \includegraphics[height=6cm,width=9cm]{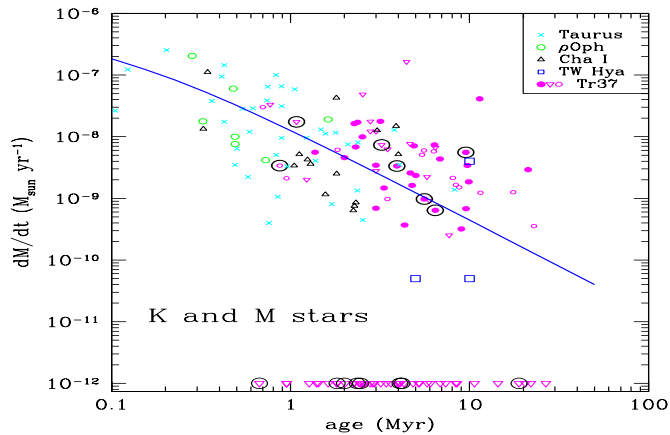} 
  \end{minipage}
  \caption{Accretion rates of K and M stars versus age in Tr 37,
  compared to the rates in other regions (Muzerolle et al. 2000, 
  APJ 535, L47) and to a viscous disk evolutionary model (Hartmann et
  al.~1998, ApJ 495, 385).  The transition objects (TO) with inner
  holes are marked with large circles.\vspace*{-1cm}}
\end{figure}

\noindent {\bf Aurora Sicilia-Aguilar} summarized how the accretion
process can be analyzed by U-band photometry and H$\alpha$
spectroscopy, helping to understand the mechanisms involved in the
disk dissipation and in the formation of inner gaps. The U-band excess
method is often preferable for K-M stars and has less systematic
uncertainties as compared to H$\alpha$, but it is also limited to
about $\dot{\rm M}\!>\!10^{-10}\rm\,M_\odot/yr$
(Sicilia-Aguilar\etal2006b\nocite{Sicilia2006b}).  Comparing the
results for intermediate-aged young clusters, the median accretion
rate in the 4-Myr-old cluster Tr 37 is $\dot{\rm
M}\!\sim\!10^{-9}\rm\,M_\odot/yr$, which is about one order of
magnitude lower than typical rates in Taurus.  About 50\% of the
transition disks\footnote{``Transition disks'' are here considered as
disks with inner opacity holes in the near IR.}  in Tr 37 are
accreting, and their accretion rates are low, but not special in
comparison to other normal disks. They are typically higher than what
is required for photoevaporation to become efficient. Observations in
Taurus (Najita\etal2007\nocite{Najita2007}) reveal similar accretion
rates for transitional disks, despite the age difference between both
clusters and the differences in the accretion rates for normal disks,
which may be an indication of common properties among transitional
disks of different ages.

Figure~2 shows that accretion rates generally decrease with age, but
there are large individual variations for stars with similar ages and
spectral types. Finally, objects with no evidence of accretion
($\dot{\rm M}\!<\!10^{-12}\rm\,M_\odot/yr$), having typically no disks
or transition disks, are found at all ages. Read more in
Sicilia-Aguilar et al. (2005, 2006a,b, 2008)
\nocite{Sicilia2005,Sicilia2006a,Sicilia2008}.

\subsubsection{3. Childhood to Adolescence: Dust and Gas Clearing in Disks}

\begin{figure}
  \begin{tabular}{cc}
  \includegraphics[height=6.5cm,angle=270]{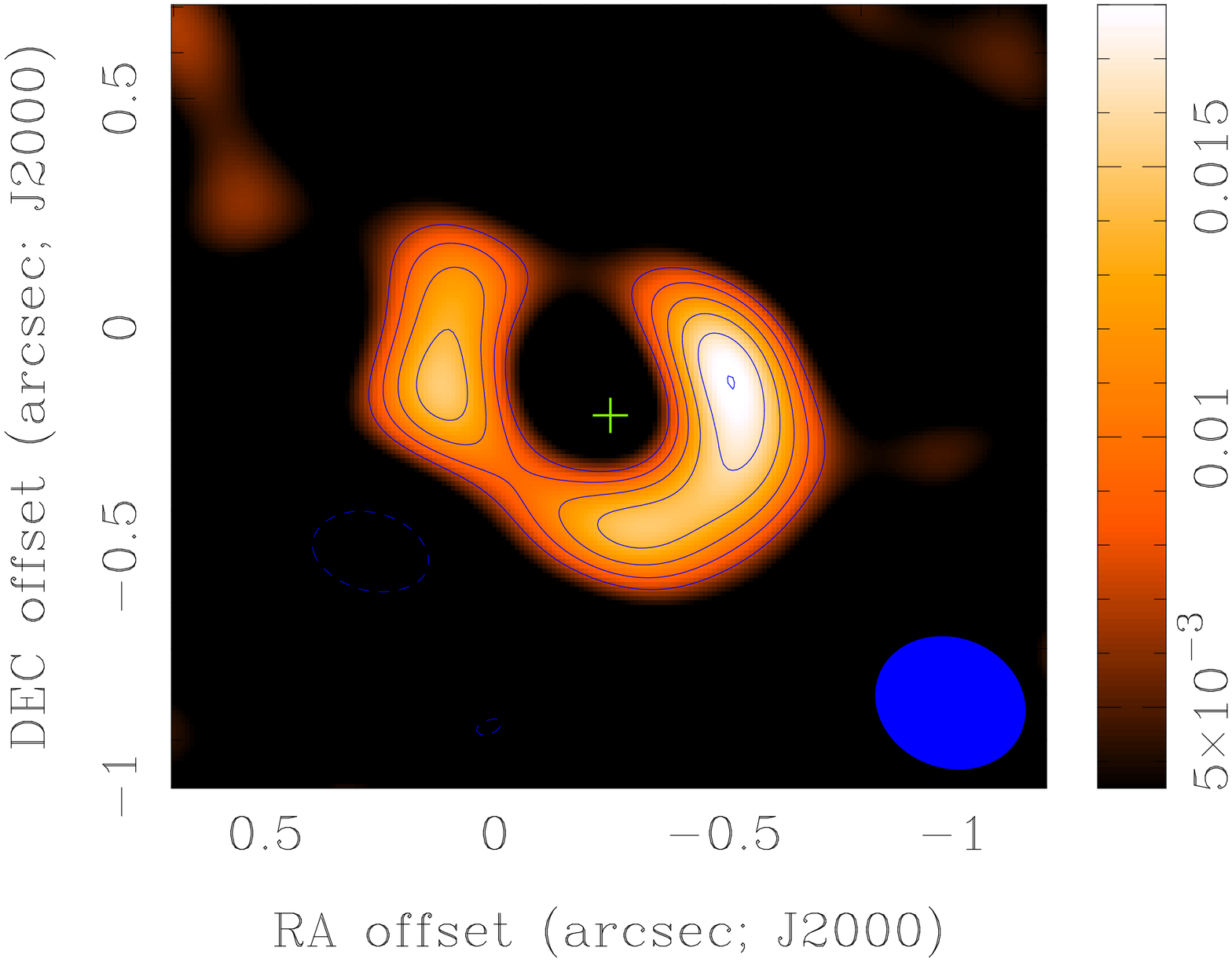} & 
  \includegraphics[height=6.5cm,angle=270]{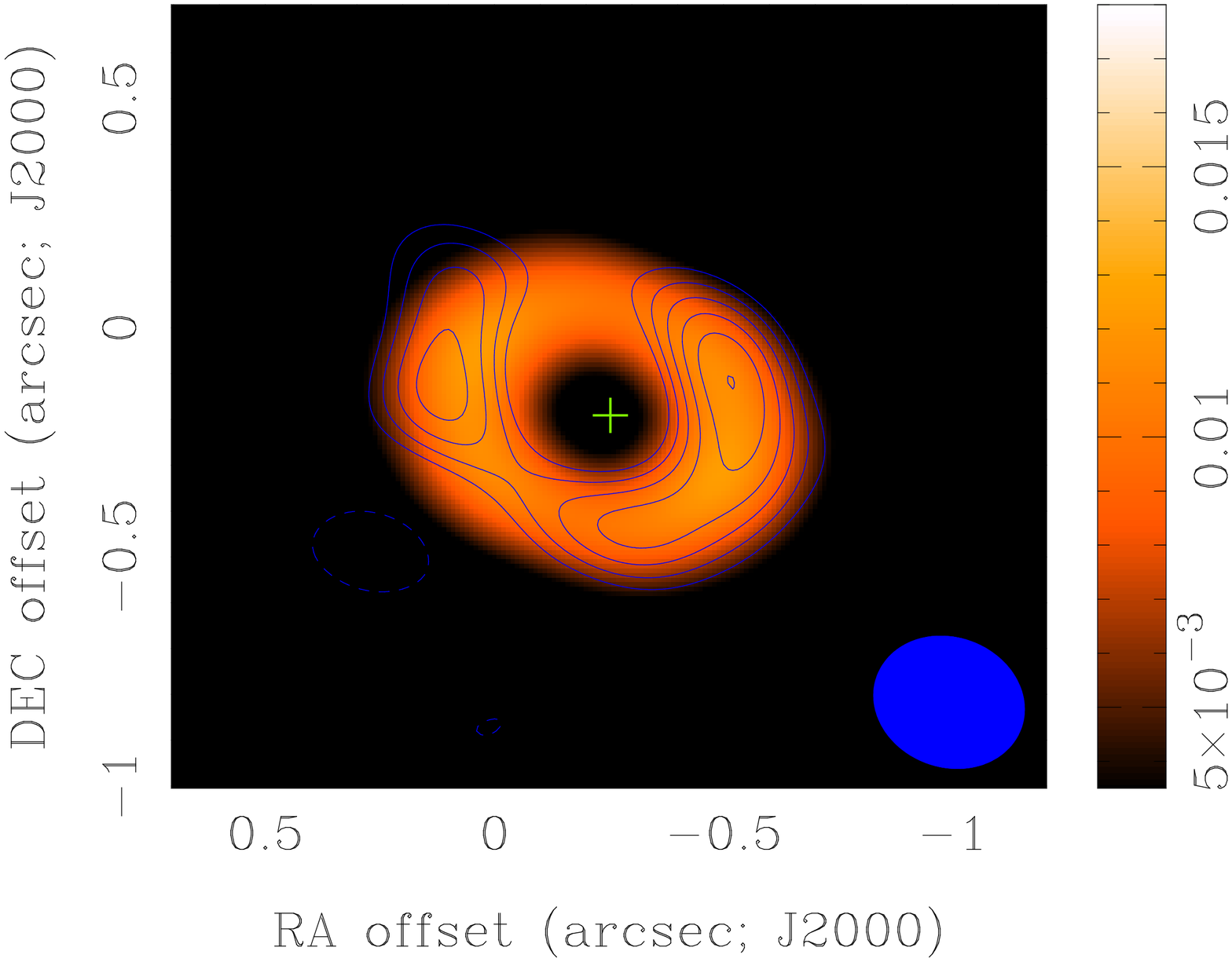}  
  \end{tabular} 
  \caption{(a) SMA 340 GHz dust continuum image of LkH$\alpha$\,330
  clearly showing an inner disk hole of approximately 40 AU radius.
  The 0\farcs28\,x\,0\farcs33 beam is plotted at the bottom right.  (b)
  Model of LkH$\alpha$\,330 overlaid with contours from the data. The
  model, based on spectrophotometry, is in good agreement with the
  data lending confidence to current interpretations of SEDs with
  significant dust emission deficits in the inner regions.}
\end{figure}

\noindent {\bf Joanna Brown} discussed different observations
of inner holes in protoplanetary disks.
Mid-infrared spectrophotometry of protoplanetary disks have revealed a
small sub-class of objects with spectral energy distributions (SEDs)
that suggest the presence of large inner gaps with low dust content,
often interpreted as a signature of young planets. New 340 GHz radio
maps (see Fig.~3) confirm this hypothesis, showing direct evidence for
the absence of dust opacity in the innermost 20-50 AU of
LkH$\alpha$\,330, SR\,21 and HD\,135344, which is in excellent
agreement with the predictions from SED modeling.

However, the SEDs are notoriously difficult to interpret as multiple
physical scenarios can result in the same SED. An alternative
explanation to the planet interaction hypothesis is that the ``gaps''
are simply a consequence of faster grain growth in the inner regions,
leading to an effective reduction of grain opacity (\eg Dullemond\plus
Dominik 2005)\nocite{Dullemond2005}. Alternatively, EUV radiation can
ionize the gas at the surface of the inner disks, which makes it
gravitationally unbound, creating true physical gaps by the action of winds 
as time progresses, even without planet interaction
(Alexander\etal 2006a,b)\nocite{Alexander2006a,Alexander2006b}.

These different scenarios should be distinguishable by gas signatures.
In fact, high resolution near IR spectra from Keck NIRSPEC and VLT CRIRES
reveal that CO gas is present well inside these ``dust gaps'' 
(Pontoppidan\etal 2008), \ie the gas distribution is different
from the dust distribution. It is noteworthy that, in general,
the gas distribution is different in different disks. See also
Brown\etal (2007,2008)\nocite{Brown2007,Brown2008}.

\subsubsection{4. Thermal and Chemical Models of Protoplanetary Disks}

\noindent {\bf Inga Kamp} summarized the present status of
thermo-chemical disk models which help to understand the location and
physical properties of the gas such as density, temperature, and
composition.
A number of 2D models of young protoplanetary disks (Kamp
\& Dullemond 2004, Jonkheid et al.\ 2004, Nomura\plus Millar 2005,
Gorti\plus Hollenbach 2007) are pioneering in combining a complex
chemical reaction network with a detailed description of the physics
and thermal balance in the disk. These models reveal hot
surface layers (see Fig.~1) where the gas lines
are already optically thick whereas the dust is still
optically thin.  





Compared to PDRs, protoplanetary disks have strong vertical density
gradients and the irradiation source, the central star, has a larger
impact and a different energy distribution than the diffuse
interstellar UV radiation field.  Young stars
often show strong X-rays either due to ongoing accretion of
disk/envelope material onto the star or due to chromospheric
activity. The presence of X-rays enhances the ionization fraction of
the disk surface (Glassgold\etal 2004, Meijerink\etal 2008), raises
molecular abundances due to efficient ion-molecule chemistry (\eg HCN;
Aikawa\plus Herbst 1999) and enhances line emission from molecular
tracers such as H$_2$ (Nomura\etal 2007).  

\begin{figure}
  \includegraphics[height=68mm,width=90mm]{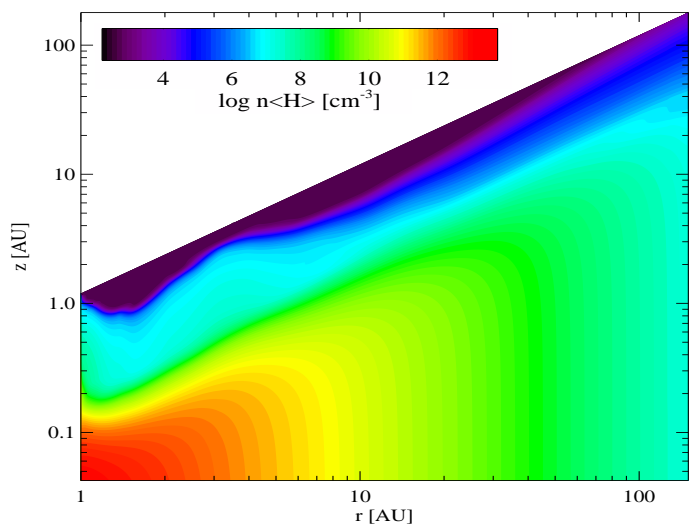} 
  \caption{Density distribution $n_{\rm\langle H\rangle}(r,z)$ in a TW
     Hya disk model with parameters $M_{\rm disk} =
     10^{-3}~$M$_\odot$, $M_\ast = 0.6~$M$_\odot$, $L_\ast =
     0.23~$L$_\odot$ and $T_{\rm eff}=4000$~K. This model includes
     the solution of hydrostatic stratification within the iteration
     of gas chemistry and energy balance, and a full 2D dust continuum
     radiative transfer (ray-based method with accelerated $\Lambda$
     iteration, Woitke, Kamp \& Thi in preparation). These
     self-consistent models do not only produce a "puffed-up" inner
     rim, but also a second less pronounced rim at
     $r\!\approx\!3-4\rm\,AU$ due to heating by absorption of stellar
     IR radiation in Fe\,{\sc ii} lines.}
\end{figure}

Future steps in the thermo-chemical disk modeling are a
self-consistent computation of the vertical disk structure, the gas
phase chemistry, and gas thermal balance.  First results from Nomura
\& Millar (2005) and Gorti \& Hollenbach (2007) contain compromises
either for the determination of the dust temperature (two-layer
approximation) or the size of the chemical network and number of
heating/cooling processes.
Also, the physical conditions in the inner few AU and the midplane
resemble more stellar atmospheres than PDRs. Hence, a series of
additional chemical reactions (three-body) and heating/cooling
processes (molecular vibrational and ro-vibrational transitions,
atomic electronic transitions, etc.) have to be added (see Fig.~4).

The fully self-consistent radiative transfer remains to-date the most
challenging problem in the thermo-chemical disk models. Line radiation
in the thermal balance is often approximated by an escape and pumping
probability formalism. More accurately, the mean radiation field
$J_\nu$ in the disk is determined by dust continuum and gas line
opacity.  Computing this mean radiation field thus requires another
loop of iteration as the gas temperature depends on $J_\nu$ (line
pumping) and the gas opacity depends on the gas composition and
temperature.


\nocite{Gorti2008}
\nocite{Kamp2005}
\nocite{Kamp2004}
\nocite{Jonkheid2004}
\nocite{Nomura2005}
\nocite{Glassgold2004}
\nocite{Meijerink2008}
\nocite{Aikawa1999}
\nocite{Nomura2007}

\subsubsection{5. Different Organic Chemistry in Disks around Sun-like
and Cool Stars}

\noindent {\bf Ilaria Pascucci} presented low-resolution {\sc Spitzer}
spectra ($\sim 7-14\mu$m) of 61 protoplanetary disks showing molecular
emission bands. It is generally assumed that the organic compounds of
the presolar nebula are representative to all protoplanetary disks,
implying that planets around different stars will all have the same bulk
composition. In a large dedicated effort with the Spitzer Space
Telescope, Pascucci\etal (2009)\nocite{Pascucci2009} show that the
abundance of simple organic molecules like C$_2$H$_2$ and HCN differ
in disks around cool stars from that in disks around sun-like stars
(see Fig.~5). Because over 80\% of stars in the galactic disk are
cooler than the Sun, this study indicates that the circumstellar
inventory of the Solar System may not be typical. HCN is likely to be
a building block of more complex organic molecules like adenine. The
sample also contains the first detections of organic molecules in the
disks around brown dwarfs. See also Pascucci\etal (2006, 2007),
Apai\etal (2005) and Luhman\etal (2007).  \nocite{Pascucci2007,
Pascucci2006, Apai2005, Luhman2007}

\begin{figure}
  \includegraphics[width=6cm,height=8cm,angle=90]{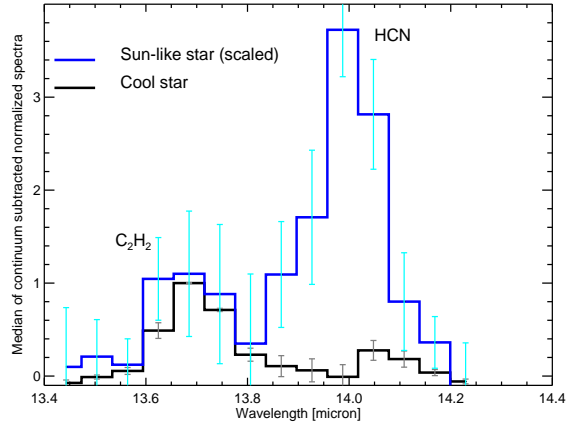} 
  \caption{Median of continuum-subtracted and normalized spectra for
  the sun-like stars (blue) and cool stars (black) samples presenting
  C$_2$H$_2$ and/or HCN emission bands.  The spectra are normalized to the
  peak of emission and scaled to match the C$_2$H$_2$ emission in the two
  samples. The error bars are the standard deviations of the
  normalized spectra. If cool stars had the same flux ratio of HCN
  versus C$_2$H$_2$ as the sun-like stars do, HCN emission would have been
  easily detected toward them.}  
\end{figure}

\subsubsection{6. Gas Dispersal and Disk Evolution}

\noindent {\bf Richard Alexander} reviewed recent developments in the
modelling of disk evolution and gas dispersal. In the "standard"
hydrodynamical picture of disk evolution, the gas is subject to
viscous transport of angular momentum and stellar photoevaporation
(Alexander et al.\ 2006a,b)\nocite{Alexander2006a,Alexander2006b}. If the
stellar EUV flux is sufficiently strong, it can ionize hydrogen, which
produces high temperatures ($T\!\approx\!10000\,$K) and small mean
molecular weights $\bar{\mu}$ at the disk surface. Outside of some critical
radius, the gas becomes thereby gravitationally unbound and a slow
($\approx\!10\rm\,km/s$) photo-evaporative wind forms. The combination
of photoevaporation and viscous evolution can form inner
holes (see Fig.~1 in Alexander\etal 2006b) and finally clears out the disk
inside-out on a relatively short timescale of about $10^{\,5}$yrs.
There are, however, a number of uncertainties in this picture, for
example the proper determination of the stellar EUV and FUV flux as
function of age and its attenuation by the accretion column. FUV
photoevaporation models need sophisticated radiative
transfer, which is difficult to couple to the hydrodynamics.


One key observation to test these models are the profiles of
the [Ne\,{\sc ii}] 12.81$\mu$m emission line. This line has been
detected by {\sc Spitzer} towards around 20 T\,Tauri stars
(Pascucci\etal 2007, Lahuis\etal 2007)\nocite{Pascucci2007,
Lahuis2007}, and is thought to arise from the upper disk
atmosphere (Herczeg\etal 2007)\nocite{Herczeg2007}.  Recent line
radiative transfer models based on photo-evaporative winds
(Alexander\etal 2008b\nocite{Alexander2008b}) predict that the line is
broad (30\,--\,40\,km/s) and double-peaked when viewed edge-on, but narrower
($\approx\!10$\,km/s) and slightly blue-shifted when viewed face-on.  This
blue-shift should be detectable with current echelle spectrographs on
8m-class telescopes, and Alexander suggested that the detection of
such a blue-shift could provide the first direct test of
photoevaporation models.  See also review article by Alexander
(2008a)\nocite{Alexander2006a}.

\subsubsection{7. Gas in Debris Disks: Clues to the Late Stages 
               of Planet Formation}

\noindent {\bf Aki Roberge} investigated the difficult case of gas in
debris disks.  These disks around nearby main sequence stars are
composed of moderate amounts of small dust grains and gas generated by
collisions between and evaporation of asteroids and comets.  The gas
component in debris disks has resisted observation for a long time and
little is known about its composition.  Nonetheless, at least some
debris disks contain gas that can be currently studied with UV/optical
absorption spectroscopy if seen edge-on.  Debris gas holds important
clues to the composition of extrasolar planetesimals during the late
stages of planet formation and the formation of terrestrial planet
atmospheres.  Figure 6 displays a highly unusual and interesting
characteristic of the gas in the Beta Pic debris disk as compared to
younger protoplanetary disks: an extreme carbon overabundance.
Details may be found in Roberge\etal (2006)
and Roberge\plus Weinberger (2008)\nocite{Roberge2006,Roberge2008}.

\begin{figure}
  \includegraphics[height=5.5cm]{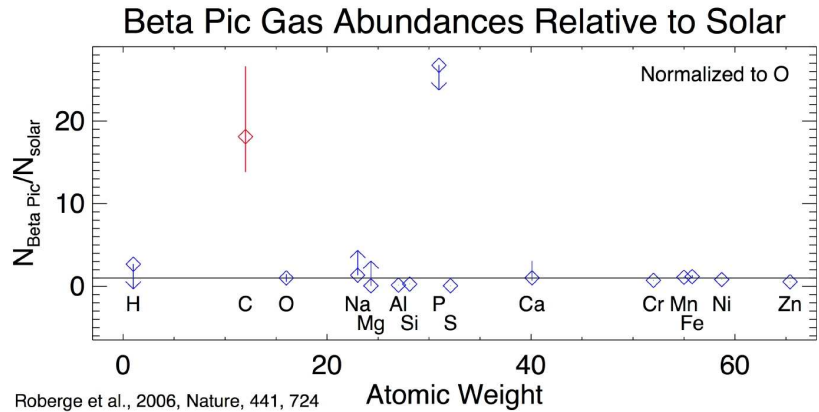} 
  \caption{The bulk composition of the Beta Pictoris circumstellar
  gas.  The midplane elemental abundances relative to solar abundances
  are plotted with diamonds.  Upper and lower limits are indicated by
  arrows.  The abundances are normalized to oxygen.  Most measured
  elements have nearly solar abundances, as does the central star.
  The exception is carbon (red diamond), which is highly
  overabundant.}
\end{figure}

\section{Conclusions}

The gas component in protoplanetary disks can be observed by emission
lines from the ultraviolet to millimeter. The structure, dynamics,
temperature structure and chemical composition of the gas in the disks
provide the initial for planet formation. The evolution of the gas
component from massive gas-rich disks to almost gas-free debris disk,
including the formation of central holes, limits the timescale for
giant planet formation, and the delivery of water and organic material
to terrestrial planets.

Spectroscopic measurements of gas in disks are generally more
challenging than broadband observations of dust both in terms of
observations and theory, but they provide complementary information on
the disk structure, in addition to unique constraints on
kinematics and chemistry. Future facilities, operating at
near-infrared to millimeter wavelengths such as {\sc Jwst, Herschel,
Sofia}, and {\sc Alma}, will open up this field and lead to dramatic
advances in our understanding of disk evolution and planet formation.




\bibliographystyle{aipproc}   


\begin{thebibliography}{9}


\bibitem[]{Aikawa1999}
Aikawa Y., Herbst E. (1999)
A\&A 351, 233

\bibitem[]{Alexander2006a} Alexander R.D., Clarke C.J., Pringle J.E. 
(2006a), MNRAS 369, 216

\bibitem[]{Alexander2006b} Alexander R.D., Clarke C.J., Pringle J.E. 
(2006b), MNRAS 369, 229

\bibitem[]{Alexander2008a} Alexander R. (2008a), 
NewAR 52, 60

\bibitem[]{Alexander2008b} Alexander R.D. (2008b), 
MNRAS (Letters), accepted, arXiv:0809.0316

\bibitem[]{Williams2007} Andrews S.M., and Williams J.P. (2007),
ApJ 659, 705

\bibitem[]{Apai2005} Apai D., Pascucci I., Bouwman J., Natta A., 
Henning Th., Dullemond C.P. (2005), Science, 310, 834

\bibitem[]{Brown2008}
Brown J.M., Blake G.A., Qi C., Dulleemond C.P., Wilner D.J. (2008),
ApJ, 675, 109

\bibitem[]{Brown2007}
Brown J.M., Blake, G.A., Dullemond C.P., Merin B., Augereau J.C.,\etal (2007), 
ApJ, 664, 107

\bibitem[]{Carr2007}
Carr J.S. (2007), 
IAU Symp.~243, "Star-Disk intereaction in Young 
Stars", ed. J.Bouvier \& I.Appenzeller, p.135

\bibitem[]{Dent2005}
Dent W.R.F., Greaves J.S., Coulson I.M. (2005), 
MNRAS, 359, 663

\bibitem[]{Dullemond2005}
Dullemond C.P., Dominik C. (2005),
A\&A 434, 971

\bibitem[]{Dullemond2007}
Dullemond C.P., Hollenbach D., Kamp I., D'Alessio P. (2007),
PPV, p.~555

\bibitem[]{Glassgold2004}
Glassgold A.E., Najita J., Igea J. (2004),
ApJ 615, 972

\bibitem[]{Gorti2008}
Gorti U., Hollenbach D. (2008),
ApJ 683, 287

\bibitem[]{Herczeg2007}
Herczeg G.J., Najita J.R., Hillenbrand L.A., Pascucci I. (2007),
ApJ 670, 509

\bibitem[]{Hughes2008}
Hughes A.M., Wilner D.J., Kamp I., Hogerheijde M.R. (2008), 
ApJ 681, 626

\bibitem[]{Jonkheid2004}
Jonkheid B., Faas F.G.A., van Zadelhoff G.-J., van Dishoeck E.F. (2004),
A\&A 428, 511

\bibitem[]{Kamp2004}
Kamp I., Dullemond C.P. (2004),
ApJ 615, 991

\bibitem[]{Kamp2005}
Kamp I., Dullemond C.P., Hogerheijde M., Enriquez J.E. (2005),
IAU Symposium 231, page 377

\bibitem[]{Lahuis2007}
Lahuis F., van Dishoeck E.F., Blake G.A.,
Evans N.J., Kessler-Silacci J.E., Pontoppidan K.M. (2007)
ApJ, 665, 492

\bibitem[]{Luhman2007} Luhman K.L., Joergens V., Lada C., Muzerolle J.,
Pascucci I., White R. (2007), PPV, p.~443 

\bibitem[]{Meijerink2008}
Meijerink R., Glassgold A.E., Najita J.R. (2008),
ApJ 676, 518

\bibitem[]{Najita2007}
Najita J.R., Strom S.E., Muzerolle J. (2007),
MNRAS 378, 369

\bibitem[]{Nomura2005}
Nomura H., Millar T.J. (2005),
A\&A 438, 923

\bibitem[]{Nomura2007}
Nomura H., Aikawa Y., Tsujimoto M., Nakagawa Y., Millar T.J. (2007),
ApJ 661, 334

\bibitem[]{Pascucci2006} Pascucci I., Gorti, U., Hollenbach D. et
al. (2006), ApJ, 651, 1177

\bibitem[]{Pascucci2007} Pascucci I., Hollenbach, D., Najita J. et
al. (2007), ApJ, 663, 383 

\bibitem[]{Pascucci2009} Pascucci I., Apai D., Luhman K., Henning Th, 
Meyer M., and Bouwman J. (2009), 
ApJ in prep.

\bibitem[]{Pontoppidan2008} Pontoppidan K.M., Blake G.A., van Dishoeck
E.F., Smette A., Ireland M.J., Brown J. (2008),
ApJ in press, arXiv:0805.3314

\bibitem[]{Qi2008}
Qi C., Wilner D.J., Aikawa Y., Blake G.A., Hogerheijde M.R. (2008), 
ApJ 681, 1396

\bibitem[]{Roberge2006} Roberge A., Feldman P.D., Weinberger A.J., 
Deleuil M., Bouret J.-C. (2006), 
Nature 441, 724

\bibitem[]{Roberge2008} Roberge A., Weinberger A.J. (2008),
ApJ 676, 509

\bibitem[]{Salyk2008} 
Salyk C., Pontoppidan K.M., Blake G.A., Lahuis F., van Dishoeck E.F.,
Evans N.J. (2008),
ApJ 676, 49S

\bibitem[]{Sicilia2006b}
Sicilia-Aguilar A., Hartmann L., F\"{u}r\'{e}sz G., Henning Th.,
Dullemond C., Brandner W. (2006b), 
AJ 132, 2135

\bibitem[]{Sicilia2008}
Sicilia-Aguilar A., Henning Th., Hartmann L. (2008), 
in prep.

\bibitem[]{Sicilia2005} Sicilia-Aguilar,
Hartmann A., Hern\'{a}ndez L., Brice\~{n}o J., Calvet N. (2005),
AJ 130, 188

\bibitem[]{Sicilia2006a} Sicilia-Aguilar A., Hartmann L., Calvet N., 
Megeath S.T., Muzerolle J., Allen L., D'Alessio P.,
Mer\'{\i}n B., Stauffer J., Young E., Lada C. (2006a), 
ApJ 638, 897

\bibitem[]{Watson2007}
Watson D.M.\etal (2007), 
Nature, 448, 1026














\end{thebibliography}



\end{document}